\documentstyle[multicol,aps,pre,twocolumn,psfig]{revtex}
\begin{document}
\twocolumn[\hsize\textwidth\columnwidth\hsize\csname@twocolumnfalse\endcsname
\title{Passive scalar intermittency in compressible flow}
\author{A.~Celani$^{1,2}$, A.~Lanotte$^{1}$,
A.~Mazzino$^{3}$ \\
\small{$^{1}$ CNRS, Observatoire de la C\^ote d'Azur, B.P. 4229,
06304 Nice Cedex 4, France.\\
\small$^2$ Dipartimento di Fisica Generale, Universit\`a di Torino,
and INFM Unit\`a di Torino Universit\`a, 
I--10126  Torino, Italy.}\\
\small$^3$ INFM--Dipartimento di Fisica, Universit\`a di Genova, I--16146
Genova, Italy.}
%\begin{document}
\draft
\date{\today}
\maketitle
\begin{abstract}
A compressible generalization of the Kraichnan model 
(Phys.~Rev.~Lett. {\bf 72}, 1016 (1994))
of passive scalar advection is considered.
The dynamical role of compressibility on the intermittency of the
scalar statistics is investigated for the direct cascade regime.
% by using a new Lagrangian method.
Simple physical arguments suggest that an 
enhanced intermittency should appear for increasing compressibility, 
due to the slowing down of 
Lagrangian trajectory separations. This is confirmed by a numerical study
of the dependence of intermittency exponents  on 
the degree of compressibility, by a Lagrangian method for 
calculating simultaneous $N$-point tracer correlations.

\end{abstract}
]
%\section{Introduction}
%\label{sec:intro}
In the last few years, much effort has been devoted to the study of 
statistical properties of scalar quantities advected by random flows with
short memory. Remarkable progress in understanding intermittency and anomalous
scaling has been achieved \cite{K94,GK95,CFKL95,SS95} for the Kraichnan 
model \cite{K94} of passive scalar advection by random, Gaussian, 
incompressible and white-in-time velocity fields. A crucial property of the 
model is that equal-time correlation functions obey closed equation of motion.
Analytical treatments are thus feasible, and the identification of a general
mechanism for intermittency has been established. Its source has been found 
in zero modes of the operators governing the Eulerian dynamics of $N$-point
correlation functions \cite{GK95,CFKL95,SS96}.\\ 
Concerning numerical studies of the Kraichnan model, efficient
Lagrangian  methods
have been recently proposed \cite{FMV98,GPZ98} and thanks to them both the 
limits
of the vanishing of intermittency corrections, for which  perturbative
predictions are available \cite{GK95,SS95}, and the non-perturbative 
region, have been successfully investigated  \cite{FMV98,FMNV98b}.

A compressible generalization of the Kraichnan model has been  recently 
proposed \cite{CKV97,GV98,AA98} and the existence of very different behaviors for
the Lagrangian trajectories, depending on the degree of compressibility, has 
been shown analytically \cite{CKV97,GV98}. 
For weak compressibility, the well-known
direct cascade of the passive scalar energy takes place. This is 
associated, from a Lagrangian point 
of view, to the explosive separation of initially close trajectories 
\cite{FMNV98b,BGK97},
a feature characterizing the direct energy cascade for the incompressible
Kraichnan model as well. On the contrary, when the compressibility is strong enough, 
particles collapse: both non intermittent inverse cascade  of tracer energy exciting large scales 
and suppression of the short-scale dissipation occur \cite{GV98}.
The relation between intermittency and compressibility is 
the main issue of the present short communication.\\
As already highlighted \cite{CKV97,GV98}, because compressibility inhibits 
the separation between Lagrangian trajectories, 
the resulting scalar transport slows down and scaling 
properties may be affected.
Our remark here is that the slowing down of Lagrangian separations plays an essential role in 
characterizing intermittency in the direct cascade regime.
This can be easily grasped from the following considerations.
In the direct cascade regime, typical trajectories are stretched, whereas  
contractions are rare and thus affect only the 
extreme tails of the pdf of scalar differences. 
Furthermore, within a Lagrangian framework, scalar correlations are
essentially governed by the time spent by particles with their
mutual distances smaller than the integral scale of the problem.
The stretching process, typical of the direct energy cascade, is thus
intermittent because contracted trajectories cause strong fluctuations
of the time needed to reach the integral scale. 
When compressibility is present, even if weakly, trapping effects are 
amplified due to the slowing 
down of Lagrangian separations. It then follows that
the dynamical role of collapsing trajectories increases for increasing
compressibility, and the same should happen for the intermittency. 
It is worth noting that the trapping mechanism, enhanced by the 
compressibility, works in the same direction as that induced
by lowering the spatial dimension $d$: it is indeed observed  perturbatively 
\cite{CFKL95} that when $d$ is reduced an increased intermittency arises, 
a fact corroborated by numerical evidences \cite{FMNV98b} comparing results of
the incompressible Kraichnan model in two and three dimensions.
These considerations will be here quantitatively supported by numerical
simulations.
\vspace{3mm}

The compressible generalization of the 
Kraichnan model is governed by the equation (for the Eulerian dynamics)
\begin{equation}
\label{fp}
\partial_t\theta(\bbox{r},t)+\bbox{v}(\bbox{r},t)\cdot\nabla\,
\theta(\bbox{r},t)=\kappa\nabla^2\theta(\bbox{r},t)+f(\bbox{r},t) ,
\end{equation}
where, as for the incompressible case, the velocity and the forcing
are zero mean, Gaussian independent processes, both homogeneous, isotropic
and white-in-time. The velocity is self-similar, with the 2-point correlation 
function:
\begin{equation}
\label{2-point-v}
\langle  v_{\alpha}(\bbox{r},t)v_{\beta}(\bbox{r}',t') \rangle =
\delta(t-t')\,\left[ d^0_{\alpha\beta} -d_{\alpha\beta}(\bbox{r}- \bbox{r}')
\right]   ,
\end{equation}
where $d_{\alpha\beta}(\bbox{r})$, the so-called {\it eddy-diffusivity}, is fixed by isotropy and scaling behavior 
along the scales:
\begin{eqnarray}
&&d_{\alpha\beta}(\bbox{r})=\nonumber\\
&&r^{\xi}\left\{\left[A+(d+\xi-1)B\right]
\delta_{\alpha\beta} + \xi \left[ A-B\right]\frac{r_{\alpha}r_{\beta}}{r^2} 
\right\} ,
\label{eddydiff}
\end{eqnarray}
where $d$ is the dimension of the space.\\
The degree of compressibility is controlled by the ratio
$\wp\equiv {\cal C}^2/{\cal S}^2$, being ${\cal S}^2\equiv 
A+(d-1)B\propto\langle(\nabla\bbox{v})^2\rangle$ and
${\cal C}^2\equiv A\propto\langle(\nabla\cdot\bbox{v})^2\rangle$,
which satisfies the inequality $0\leq \wp \leq 1$.\\
The statistics of the forcing term is defined by the 2-point correlation 
function
\begin{equation}
\label{2-point-f}
\langle  f(\bbox{r},t)f(\bbox{r}',t') \rangle =
\delta(t-t')\,\chi(|\bbox{r}- \bbox{r}'|)  ,
\end{equation}
where $\chi$ is chosen nearly constant for distance $|\bbox{r}- \bbox{r}'|$
smaller than the integral scale $L$ and rapidly decreasing for $r \gg L$.\\
It is worth remarking that equation (\ref{fp}) physically describes the 
evolution of a tracer, that is a quantity which is conserved
along the Lagrangian trajectories in absence of diffusivity and forcing. 
To characterize the advection of a density, one should consider
the equation
\begin{equation}
\label{density}
\partial_t\rho(\bbox{r},t)+\nabla \cdot \left( \bbox{v}(\bbox{r},t)
\rho(\bbox{r},t) \right) =\kappa\nabla^2\rho(\bbox{r},t)+f(\bbox{r},t) ,
\end{equation}
which in the ideal case ($\kappa=0$, $f=0$) enjoys the conservation 
of the total mass. 
The density advection equation
has also a  wide realm of physical applications and should  deserve
 a detailed study in its own, as well as a specific numerical approach. 
Hereafter we shall limit ourselves 
to the case of tracer advection
ruled by (\ref{fp}).

Exploiting the $\delta$-correlation in time, equations for the even scalar 
correlations (odd correlations being trivially zero) in the stationary state,
can be deduced \cite{GK96}; for the generic $N$-point correlation function $C_N^{\theta}\equiv\langle \theta(r_1)\cdots\theta(r_N) \rangle$ the expression reads:
\begin{equation}
{\cal M}_N\,C_N^{\theta}=
\sum_{i<j} \chi\left( \frac{r_{ij}}{L}\right)
\langle \theta(r_1)
\smash{\mathop{\dots\dots}_{\hat{i}\ \ \hat{j}}}
\theta(r_N) \rangle \;, 
\label{eqclosed}
\end{equation}
with $r_{ij}\equiv r_i-r_j$, and ${\cal M}_N$ is the differential operator
given by:
\begin{eqnarray}
&&{\cal M}_N=\nonumber\\
&& \sum_{1\leq n < m \leq N} d_{\alpha\beta}(\bbox{r}_n - \bbox{r}_m)
\nabla_{r_{n{\alpha}}} \nabla_{r_{m{\beta}}} - \kappa\sum_{1\leq n\leq N}
\nabla^2_{r_n}  .
\label{emme}
\end{eqnarray}

As for the incompressible case, this model has a Gaussian limit for $\xi\to 0$,
and the perturbative expansion at small $\xi$'s can be done as in 
Ref.~\cite{GK95}. Accordingly, the calculation performed in the weakly 
compressible case (i.e. $\wp < d/\xi^2$) 
corresponding to the direct cascade regime leads
(see Ref.~\cite{GV98})
to the expression for the intermittent correction $\Delta_N^{\theta}$, to the 
normal scaling exponent $(2-\xi)N/2$ of the N-point structure function
 $S_N^{\theta}(r) = \langle [\theta(\bbox{r}) -\theta(\bbox{0})]^N\rangle
\propto r^{(2-\xi)N/2-\Delta_N^{\theta}}$; namely:
\begin{equation}
\Delta_N^{\theta}=\frac{N(N-2)(1+2\wp)}{2(d-2)} \xi + O(\xi^2)  .
\label{perturb}
\end{equation}
The perturbative approach gives thus a first clue that compressibility 
works to enhance intermittent corrections. 
We are however interested in checking that this is a general and robust 
feature associated to compressibility and thus that it is
present for generic $\xi$. 
This problem is not accessible by 
perturbative techniques; numerical methods are generally needed 
to investigate it.
With this purpose in mind, we have developed a new 
Lagrangian numerical method 
(a different viewpoint with respect to the one in Ref.~\cite{FMV98}), 
where the strategy is now
formulated in terms of a {\em first exit time} problem \cite{gaw}. 
%The main advantage is
%that no free-boundary conditions have to be imposed for asymptotic separations.
%This implies a reduction of computational costs,
%and thus an easier investigation.\\

%\section{The Lagrangian method}
%\label{sec:method}
%\vspace{0.15 cm}
%According to the zero-modes picture of passive scalar intermittency, 
%the anomalous scaling of the structure function 
%of order $N$
%is related to the existence of non-trivial scaling solutions of
%the homogeneous equation ${\cal M}_{N} {\cal Z} = 0$ \cite{GK95,BGK97}.
%As a matter of fact, the analytic computation of such solutions
%is -- in general -- a daunting task.
%On the other hand, it is possible to devise a numerical method 
%which allows a precise determination of the anomalies.

The method consists in the Montecarlo simulation of 
Lagrangian trajectories according to the stochastic differential equation
\begin{equation}
\dot {\bbox{r}}_n = \bbox{v}(\bbox{r}_n,t)+\sqrt{2\kappa}\dot{w}_n \;,
\end{equation}
where the $w_n$ are independent Wiener processes.
The evolution of the probability $ P_{N}(t,\bbox{x}|t_0,\bbox{x}_0) $ that
the $N$ Lagrangian tracers have a configuration
 $\bbox{x}=(\bbox{r}_1,\ldots,\bbox{r}_N)$  at time
$t$ given their initial configuration  $\bbox{x}_0$ at time 
$t_0$ is ruled by
the Fokker-Planck equation
\begin{equation}
\frac{\partial}{\partial t} P_{N}(t,\bbox{x}|t_0,\bbox{x}_0)+
{\cal M}^{\star}_{N}(\bbox{x}) P_{N}(t,\bbox{x}|t_0,\bbox{x}_0) = 0 \;, 
\label{2.1}
\end{equation}
where the operator ${\cal M}^{\star}_{N}$ is the adjoint of (\ref{emme}).
As a consequence of (\ref{2.1}) the probability
obeys also the backward Kolmogorov  equation
\begin{equation}
\frac{\partial}{\partial t_0} P_{N}(t,\bbox{x}|t_0,\bbox{x}_0) +
{\cal M}_{N}(\bbox{x}_0) P_{N}(t,\bbox{x}|t_0,\bbox{x}_0) = 0 \;.
\label{2.11}
\end{equation}
We now introduce the Green function
\begin{equation}
G(\bbox{x},\bbox{x}_0)=
\int_{t_0}^{\infty} dt \; P_N(t,\bbox{x}|t_0,\bbox{x}_0)\;,
\label{eq:2.2}
\end{equation}
which enjoys the following properties
\begin{eqnarray}
{\cal M}^{\star}_{N}(\bbox{x}) G(\bbox{x},\bbox{x}_0) &=& 
-\delta(\bbox{x}-\bbox{x}_0) \; ,\\
{\cal M}_{N}(\bbox{x}_0) G(\bbox{x},\bbox{x}_0) &=& 
-\delta(\bbox{x}-\bbox{x}_0) \; .
\label{eq:2.21}
\end{eqnarray}
Let us define the characteristic size of a configuration of $N$ particles
as $R(\bbox{x})
=[(\sum_{i<j} |\bbox{r}_i-\bbox{r}_j|^2)/(N(N-1)/2)]^{1/2}$ .
We now impose Dirichlet (absorbing) boundary conditions
at $R(\bbox{x})=L \gg R(\bbox{x}_0)$, 
and compute numerically the first exit time from 
the volume of configuration space limited by the boundary, which is 
expressed in terms of the Green function as (see e.g. \cite{Risken})
\begin{equation}
T_L(\bbox{x}_0)=\int_{R(x)<L} dx\; G(\bbox{x},\bbox{x}_0).
\end{equation}
A trivial consequence of the property (\ref{eq:2.21}) is that
\begin{equation}
{\cal M}_{N}(\bbox{x}_0) T_L(\bbox{x}_0) = -1, 
\end{equation}
an equation whose
structure resembles that of (\ref{eqclosed}); indeed we can
 conclude, similarly to what happens for correlation functions 
(e.g.\cite{GK95,CFKL95}),
 that $T_L(\bbox{x}_0)$ must amount to the sum of an
inhomogeneous solution plus a linear combination of
zero modes $f_j$ of the operator ${\cal M}_{N}$:
\begin{equation}
T_L(\bbox{x}_0)=\sum_j C_j L^{\gamma-\sigma_j} f_j(\bbox{x}_0)
+ \mbox{inhomog. term}\;,
\label{2.5}
\end{equation}
where the explicit dependence on $L$ has been extracted
taking advantage of the scaling properties of ${\cal M}_{N}$,
$\sigma_j$ is the scaling exponent of the zero mode $f_j$ and 
$C_j$ is a constant independent of $L$.
Among the non trivial zero-modes $f_j$,
only the functions which depend on all
the coordinates can contribute to the $N$-th order structure function.
We would like to extract this contribution leaving aside all the others:
it is easy to realize that this result can be achieved
performing a linear combination of the
exit times with different initial conditions.
This operation will remove also the inhomogeneous term.  
If we denote with $\nabla_i(\bbox{\rho})$ the operator acting on the functions
of $N$-particles coordinates as 
$\nabla_i(\bbox{\rho})F(\bbox{r}_1,\ldots,\bbox{r}_i,\ldots,\bbox{r}_N)
=F(\bbox{r}_1,\ldots,\bbox{r}_i+\bbox{\rho},\ldots,\bbox{r}_N)-
F(\bbox{r}_1,\ldots,\bbox{r}_i,\ldots,\bbox{r}_N)$
we will have 
\begin{equation}
\Sigma_N(L)=\prod_i \nabla_i(\bbox{\rho}) T_L(\bbox{x}_0) \propto 
L^{\gamma-\zeta_N}
\label{2.6} 
\end{equation}
where $\zeta_N=(2-\xi)N/2-\Delta^{\theta}_N$ is the scaling exponent of the
structure function $S_N^{\theta}(r)\sim r^{\zeta_N}$.
Whenever $\bbox{x}_0=\bbox{0}$, due to the simmetry of the
$f_j$'s under exchanges of particles coordinates, 
the expression for $\Sigma_N(L)$ takes a
simple form, which,
for example, for $N=4$ reads as
%\begin{eqnarray}
%\Sigma_4(L)&=&
%2\, T_L(\bbox{0},\bbox{0},\bbox{0},\bbox{0})-
%8\, T_L(\bbox{\rho},\bbox{0},\bbox{0},\bbox{0})+ \nonumber \\
% & & 6\, T_L(\bbox{\rho},\bbox{\rho},\bbox{0},\bbox{0}) \; .
%\label{2.7}
%\end{eqnarray}
$
\Sigma_4(L)=
2\, T_L(\bbox{0},\bbox{0},\bbox{0},\bbox{0})-
8\, T_L(\bbox{\rho},\bbox{0},\bbox{0},\bbox{0})+ 
6\, T_L(\bbox{\rho},\bbox{\rho},\bbox{0},\bbox{0})
$.

Summarizing: the numerical method consists 
in the Monte-Carlo simulation of Lagrangian trajectories
of $N$ particles advected by a rapidly changing velocity field,
according to the Fokker-Planck equation (\ref{2.1});
average first exit times outside a volume of size $L$ are computed 
for different arrangements of the initial conditions, and then
linearly combined according to (\ref{2.6}) in order to extract
the scaling exponent $\zeta_N$.

As a final remark, the numerical method here employed
can be viewed as a merging of the two Lagrangian methods  
introduced by Frisch, Mazzino and Vergassola in Ref.~\cite{FMV98} 
and by Gat and Zeitak in Ref.~\cite{GZ97}.
Namely, it borrows form the first one the idea of subtracting 
exit times of different initial conditions to extract the 
only zero mode that contributes to the structure functions,
while inherits from the second the spirit of working with 
particle configurations (shapes). 
The advantages of the present method with respect to \cite{FMV98}
mainly reside in the evaluation of first exit times rather than
of residence times, a fact which substantially reduces the computational cost.

We present the numerical results obtained for the scaling of the 
fourth-order structure function  
$S_4(r;L)\equiv \langle(\theta(\bbox{r}) - \theta(\bbox{0}))^4\rangle$ in  three dimensions. As previously mentioned, when the dimension $d$ of the space is lowered fluctuations increase and  as a consequence the number of realizations needed to have a clean scaling grows as well; the addition of compressibility further enhances this effect.
For the first numerical experiments with the new method,  we have thus opted 
for $d=3$.\\
The method has been tested performing the analysis 
of the incompressible 
limit $\wp=0$ for different values of $\xi$: 
the anomaly $\Delta_4^{\theta}=2\zeta_2 - \zeta_4 $
has always been found to be compatible with the results presented in 
Refs.~\cite{FMV98,FMNV98b}. The computation of
$\Sigma_2(L)$ -- which can be evaluated
analitically -- has provided another stringent test for this method.\\ 
Varying the degree of compressibility $\wp$, we have studied 
in the direct cascade regime the connection between 
the slowing down of Lagrangian trajectories and intermittency at the two 
distinct values
$\xi=0.75$ and $\xi=1.1$. 
Notice that for these two values of $\xi$, the condition ($\wp < d/\xi^2$) 
for the direct cascade 
of energy to take place \cite{GV98} is verified for 
the entire range of values $0\leq \wp \leq 1$ of the compressibility.
Different motivations account for this choice; 
first of all
we avoided the region of $\xi$ close to $0$ 
($\gamma\rightarrow 2$) where capturing the subdominant anomalous 
exponents is numerically expensive, and furthermore the results are
known from perturbative expansion. Second,
when  $\xi$ is close to $2$ ($\gamma \rightarrow 0$) non local 
effects are very strong and 
the range of values of $\wp$ (i.e. $\wp < d/\xi^2$)
pertaining to the direct cascade is narrower. 

\begin{figure}
\centerline{\psfig{file=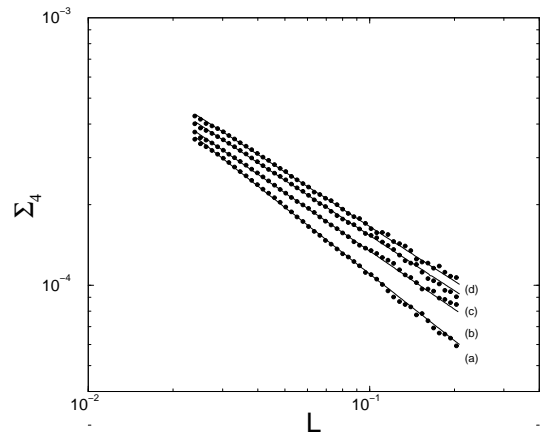,width=7cm}}
%\vspace{1mm}
\caption{A log-log plot of $\Sigma_4(L)$ for
$\xi=0.75$. (a): $\wp=0$; (b): $\wp=0.25$; (c): $\wp=0.5$; (d): $\wp=0.75$. 
Separation $\rho=2.7\times 10^{-2}$, 
diffusivity $\kappa=2.3\times 10^{-5}$, 
number of realizations ranging from $20\times 10^6$
(case (a))  to $30\times 10^6$ (case (d)).
Solid lines represent the best fit power laws.}
\label{fig1}
\end{figure}

\begin{figure}
\centerline{\psfig{file=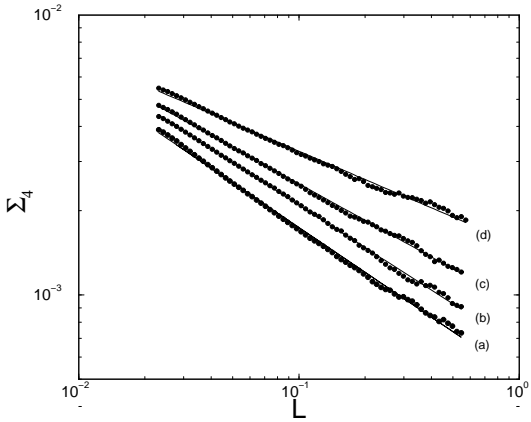,width=7cm}}
%\vspace{1mm}
\caption{As in Fig.~1, for $\xi=1.1$ and diffusivity $\kappa=2.5\times 
10^{-3}$.}
\label{fig2}
\end{figure}

In Figs.~\ref{fig1} and \ref{fig2} are shown the behavior of $\Sigma_4(L)$
for the two values of $\xi$ under consideration and for different values of $\wp$,
which all display a fairly good power law scaling.
According to the relation
(\ref{2.6}) the scaling exponent is 
$\gamma-\zeta_4= -\gamma + \Delta_4^{\theta}$, so that the curves become 
flatter and flatter as the anomaly grows . 
It is thus evident from our results that when compressibility increases, 
the intermittent correction to the normal scaling grows as well. 

%%%%%%%%%%%%%%%%%%%%%%%%%%

Notice that ratio between $\Sigma_4$
and the dominant contribution to each term of the sum
 scales as $L^{-\zeta_4}$. As a 
consequence, small values of $\xi$ (which correspond to
large values of $\zeta_4$) require 
a larger amount of statistics to 
make the subdominant  contribution emerge. 
This is the reason for which the scaling region
for $\xi=0.75$  is smaller than that for $\xi=1.1$.

\begin{figure}
\centerline{\psfig{file=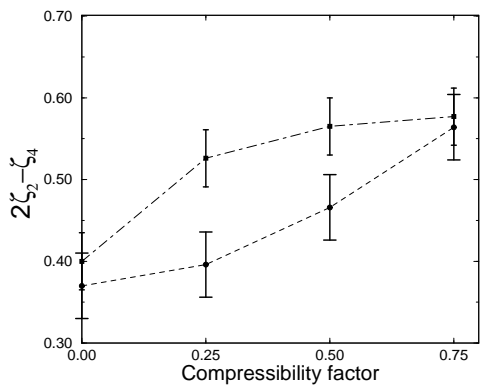,width=7cm}}
%\vspace{1mm}
\caption{The anomaly  $2\zeta_2 - \zeta_4$ for the fourth-order structure function, for $\xi=0.75$ 
(squares joined by a dot-dashed line) and $\xi=1.1$ (circles joined by a dashed line).}
\label{fig3}
\end{figure}

Finally, our results are summarized in Fig.~\ref{fig3} which shows
the  anomaly   $2\zeta_2 - \zeta_4$ {\it vs} the 
compressibility factor $\wp$ for $\xi=0.75$ 
(squares joined by a dot-dashed line) and $\xi=1.1$
(circles joined by a dashed line).
As in Ref.~\cite{FMV98}, the error
bars are obtained
by analyzing the fluctuations of local scaling exponents over octave
ratios of values for $L$, a method which gives a very conservative estimate
of the errors. The effectiveness of the first exit time computation is somehow
balanced by the need of a huge number of realizations to achieve
a satisfactory statistical convergence. This drawback is particularly
visible for large $L$, where the signal is rather noisy.

In conclusion, we have shown in the context of the Kraichnan compressible 
model 
that there is a tight relationship between intermittency of passive scalar 
statistics and compressibility of the advecting velocity field. This result 
can be easily understood from the Lagrangian viewpoint. 
Intermittency arises whenever the particles experience long 
periods of inhibited separation: since compressible flows are characterized
by the presence of trapping regions, an enhancement of intermittency can be 
reasonably expected. The validity of this
argument has been assessed by means of a numerical Lagrangian method.

We acknowledge innumerable discussions
on the subject matter with  M.~Vergassola.
Simulations were performed in the framework
of the SIVAM project of the Observatoire de la C\^ote d'Azur.  Part of
them were performed using the computing facilities of CINECA.

\end{document}